# Realizing anomalous Floquet non-Abelian band topology in photonic scattering networks

Yuze Hu[1,2,†], Mingyu Tong[1,3,4,5,†], Tian Jiang[2], Shuxing Yang[1,3,4,5], Ning Han[6], Fujia Chen[1,3,4,5], Li Zhang[1,3,4,5], Rui Zhao[1,3,4,5], Qiaolu Chen[1,3], Hongsheng Chen[1,3,4,5,*], F. Nur Ünal[7,*], Robert-Jan Slager[8,*], Yihao Yang[1,3,4,5,*]

[1] Interdisciplinary Center for Quantum Information, State Key Laboratory of Extreme Photonics and Instrumentation, ZJU-Hangzhou Global Scientific and Technological Innovation Center, Zhejiang University, Hangzhou 310027, China.
[2] College of Science, National University of Defense Technology, Changsha, 410073, China
[3] International Joint Innovation Center, The Electromagnetics Academy at Zhejiang University, Zhejiang University, Haining 314400, China.
[4] Key Lab. of Advanced Micro/Nano Electronic Devices & Smart Systems of Zhejiang, Jinhua Institute of Zhejiang University, Zhejiang University, Jinhua 321099, China.
[5] Shaoxing Institute of Zhejiang University, Zhejiang University, Shaoxing 312000, China.
[6] College of Optical and Electronic Technology, China Jiliang University, Hangzhou 310018, China.
[7] School of Physics and Astronomy, University of Birmingham, Edgbaston, Birmingham B15 2TT, United Kingdom.
[8] Department of Physics and Astronomy, University of Manchester, Oxford Road, Manchester M13 9PL, United Kingdom

[†] These authors contributed equally to this work.
[*] Corresponding author. Email:
yangyihao@zju.edu.cn (Y. Y.);
robert-jan.slager@manchester.ac.uk (R.-J. S.);
f.unal@bham.ac.uk (F. N. Ü.);
hansomchen@zju.edu.cn (H. C.)

## Abstract

The concept of multi-gap topology has recently been shown to give rise to uncharted phases beyond conventional single-gap classifications. These phases relate to band nodes with non-Abelian quaternion charges and momentum-space braiding processes characterized by new invariants such as paradigmatic Euler class, phenomena that intrinsically require at least two spatial dimensions. Extending such phases into the non-equilibrium regime is predicted to unlock even richer multi-gap topologies beyond static settings, yet their experimental realization has remained elusive due to the stringent requirements on dimensionality, symmetry, and dynamical control. Here, we theoretically demonstrate and, for the first time, experimentally realize two-dimensional (2D) Floquet non-Abelian band topology in photonic scattering networks. Within this platform, we uncover a sequence of topological phenomena unique to 2D multi-gap systems far from equilibrium, including anomalous multi-gap phases interconnected by band nodes, Floquet Euler transfer, gapped phases with anomalous Dirac string configurations, and Floquet-induced non-Abelian braiding of band nodes. In addition, we observe Floquet-periodic anomalous edge states across multiple gaps, providing experimental signatures of these sought-after 2D multi-gap Floquet topological phases. Our results establish photonic scattering networks as a practical and versatile route to non-Abelian Floquet systems, opening avenues for dynamical topological physics with braiding capability and robust photonic functionalities.

**Introduction**

The topology of band structures is usually understood one gap at a time. Bands around a single isolated gap can for example harbor a Chern number[1], a time reversal symmetry protected Kane-Mele invariant[2], or a Berry phase[3], with these quantities faithfully predicting the existence of chiral or polarization-type edge states[4–6]. However, when several bands (gaps) are considered together, new topological structures can emerge[7–16]. Specifically, recent discoveries have shown multi-gap topological invariants that cannot be expressed in terms of the well-established single-gap classifications[9,10,15,17,18]. In addition, they intricately relate to non-Abelian processes in momentum space: band nodes behave as objects that carry (generalized) quaternion-valued charges, and their motion in momentum space can lead to braiding, recombination, and topological degeneracies across multiple energy gaps[7–9,16,19]. These braiding effects require at least two spatial dimensions, and they have no analog in one-dimensional systems or in conventional single-gap topologies.

Beyond static settings, non-equilibrium platforms, such as periodically driven systems or unitary scattering networks whose step-wise propagation naturally realizes a Floquet evolution operator, unlock topological behaviors generically inaccessible in equilibrium[20,21,21–24]. A key feature of these systems is that quasi-energy is defined modulo $2\pi$, causing the spectrum to "wrap around" itself[25–33]. This periodicity fundamentally reshapes how band nodes interact: a node exiting the top of the quasi-energy zone re-enters from the bottom, allowing previously disconnected gaps to become linked. As a result, topological charges can migrate across the quasi-energy landscape, enabling anomalous phases without any static counterpart[20,21,33].

These possibilities have been recently predicted to culminate in new types of phases in multi-gap context, including Floquet Euler phases, anomalous Dirac-string configurations, and non-Abelian node braiding, which are fundamentally distinct from static or Abelian paradigms[14,29,34–36]. Despite these compelling predictions, Floquet induced braiding and non-Abelian phases in 2+1 dimensions has not been achieved in experiments so far, mainly because constructing a fully tunable multi-band Floquet platform with controlled degeneracies and symmetry constraints has remained technically inaccessible. Here we stress the crucial aspect of dimensionality as one-dimensional Floquet multiband systems have clarified important limitations: while Zak phases remain meaningful, true non-Abelian braiding does not occur in one dimension where a complete and faithful bulk–boundary correspondence is still subject of active research efforts[29,34–36].

We here overcome these fundamental challenges by theoretically demonstrating and experimentally realizing Floquet non-Abelian topology in a two-dimensional (2D) photonic scattering network, directly observing anomalous edge states across multiple quasi-energy gaps. The platform is built from a Kagome lattice of magnetically biased three-port circulators interconnected by low-loss waveguides. Although structurally simple, the network is engineered to break time-reversal symmetry ($\mathcal{T}$) and two-fold rotation ($C_{2z}$) individually while preserving their product $C_{2z}\mathcal{T}$—a symmetry condition that forces all Floquet eigenmodes to remain real and renders non-Abelian frame charges well defined[7–16,37]. Within this framework, we uncover a sequence of Floquet phases that reveal the richness of the multi-gap landscape, as shown in Fig. 1. At a critical point, the system enters an anomalous multi-gap phase in which Dirac nodes interlink all quasi-energy bands, enabling a Floquet-driven transfer of Euler-class invariants between distinct band subspaces. Beyond this transition, the network enters a phase characterized by anomalous Dirac strings, which opens gaps between two three-band branches. Within this gap, non-trivial configurations of the Dirac strings are stabilized, delineating the topological nature of the phase. Both phases arise from non-Abelian braiding of band nodes, activated by the periodic drive and responsible for reorganizing topological charges across the spectrum. Crucially, we directly observe the predicted Floquet edge states—both anomalous and antichiral—in experiment, providing clear evidence for the underlying multi-gap topology and establishing our platform as a powerful route for exploring non-Abelian band topologies far from equilibrium.

**Main Text**

**2D Floquet non-Abelian Topology in photonic scattering networks**

We present a non-reciprocal photonic network platform exhibiting unitary wave scattering, as shown in Fig. 1a. The architecture, consisting of a periodic Kagome lattice, is modelled by identical three-port non-reciprocal scattering nodes, and lossless bidirectional waveguiding channels (Fig. 1b). A feature of the unit cell design is a combined $C_{2z}\mathcal{T}$ symmetry, while the individual scattering elements introduces three-fold rotational symmetry ($C_3$) (Fig. 1b). This symmetry emerges from the specific coupling within the unit cell, which pairs two such $C_3$-symmetric scatterers subjected to mutually opposing magnetic flux biases with an inter-element phase shift $\varphi$. The fundamental scattering process at each node is characterized by a unitary $3 \times 3$ matrix $S_0$, which is inherently asymmetric due to non-reciprocity. Remarkably, this matrix is fully determined by only two independent angular parameters, $\xi$ and $\eta$, as illustrated by its general parametrization in Fig. 2e. To model wave transport across the periodic structure as a Bloch eigenproblem, this equation simultaneously accounts for the coherent scattering events at the lattice nodes within a wavevector $\boldsymbol{k}$-dependent $6 \times 6$ unitary scattering matrix $S(\boldsymbol{k})$, and the phase delay $\varphi$ accumulated during propagation along the bidirectional links[25,38]. The resulting network's wave-transport characteristics are described by:

$$S(\boldsymbol{k})|c(\boldsymbol{k})\rangle = e^{-i\varphi}|c(\boldsymbol{k})\rangle. \tag{1}$$

Formally, the wave dynamics within this non-reciprocal scattering network are governed by a unitary $6 \times 6$ matrix $S(\boldsymbol{k})$ (see Supplementary Information). This description maps directly onto a Floquet eigenvalue problem, analogous to that of a periodically driven lattice. Within this description, the angular variable $\varphi \in (-\pi, \pi]$ assumes the role of quasi-energy. Distinct from anomalous Floquet Chern insulators[24], our design employs an antiparallel flux configuration within the unit cell's two circulators. This specific 2D Kagome network architecture breaks both $\mathcal{T}$-symmetry and $C_{2z}$, while preserving the combined $C_{2z}\mathcal{T}$ symmetry. This preservation has a profound consequence: the Floquet eigenstates of Eq. 1 acquire real values. Consequently, the non-Abelian frame charges characterizing the energy bands are well-defined[9], enabling the emergence of novel non-Abelian topological phases that uniquely leverage the combined advantages of engineered non-reciprocity and Floquet driving.

Of particular relevance to establishing non-Abelian band topology, the photonic scattering network platform establishes a highly adaptable and functionally rich environment for investigating anomalous non-Abelian Floquet topological phases. This system uniquely accommodates the realization of several distinct non-equilibrium states, fundamentally inaccessible in static regimes. The gapless anomalous multi-gap phase where all Floquet bands are connected by band nodes, facilitating the transfer of Euler class invariants between different band subspaces, as shown in Fig. 1c; the anomalous Dirac string phase, a gapped state between the two three-band subspaces characterized by a non-trivial Dirac string configuration within gaps, as shown in Fig. 1d; and the underlying Floquet-induced non-Abelian braiding processes of band nodes, which govern Euler transfer between different topological phases, as shown in Fig. 1e. The capacity of our platform to host such diverse phenomena underscores its utility in probing anomalous Floquet multi-gap topology and provides a direct pathway for observing associated boundary signatures, thereby offering a versatile testbed for exploring exotic non-Abelian physics in driven systems.

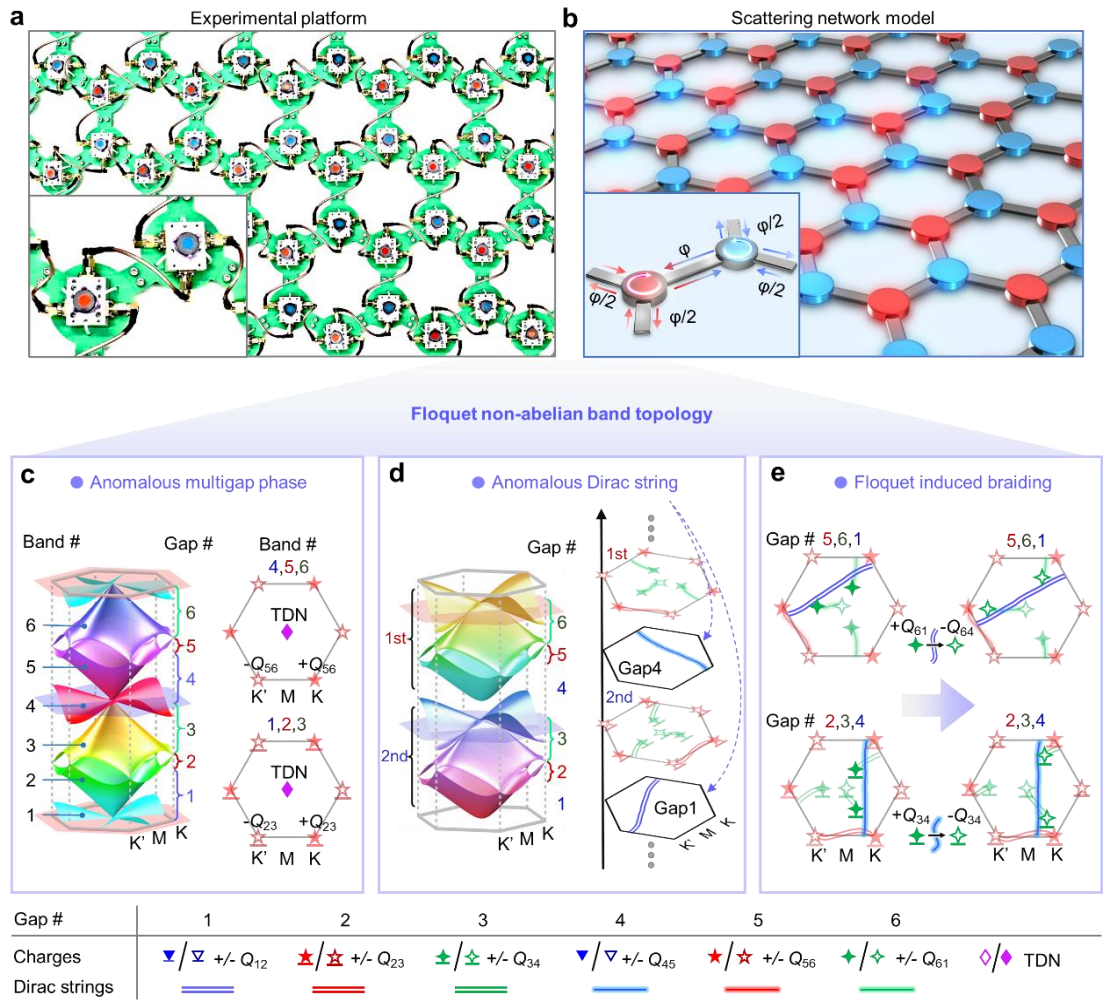

**Fig. 1 | 2D Floquet non-Abelian band topology in photonic scattering networks. a**, **b**, Experimental platform (a) and theoretical model (b) of a non-Abelian photonic scattering network.

The non-reciprocity scattering is realized by staggered magnetic flux in three-port circulators and the nearest-neighbor hopping between connected lines is governed by asymmetric unitary scattering matrices. Insets show one-unit cell with scattering elements connected by coaxial transmission lines that induce a phase delay (φ). The cell scattering operator generates Floquet bands with quasi-energy φ. **c**, Anomalous multigap phase of energy dispersions and band nodes when experiencing a Floquet Euler transfer. The six bands are connected together by triple degeneracy nodes. **d**, Anomalous Dirac strings are formed in the gaps between two similar branches of three bands. **e**, Braiding of the Dirac nodes is facilitated by crossing the anomalous Dirac strings owing to the Floquet periodicity.

## Anomalous multigap phase and Floquet Euler transfer

In $C_{2z}\mathcal{T}$-symmetric systems, the Euler class, $\epsilon$, quantifies the topology of isolated two-band subspaces, classifying the braiding and movement of band nodes[8,13,14]. Specifically, considering the band subspace $(n, n+1)$ spanned by Bloch states $|u_n(\mathbf{k})\rangle$ and $|u_{n+1}(\mathbf{k})\rangle$ where $n$ is the gap number, the reality condition can be employed to define an Euler form for these bands[8,10,11,14]:

$$\text{Eu} = \langle \partial_{k_x} u_n(\mathbf{k}) | \partial_{k_y} u_{n+1}(\mathbf{k}) \rangle - \langle \partial_{k_y} u_n(\mathbf{k}) | \partial_{k_x} u_{n+1}(\mathbf{k}) \rangle \tag{2}$$

and an Euler connection, $A = \langle u_n(\mathbf{k}) | \nabla u_{n+1}(\mathbf{k}) \rangle \cdot d\mathbf{k}$. The patch Euler class is evaluated by integrating over a patch $D$ in the Brillouin zone that excludes the nodes form neighboring gaps

$$\epsilon_n[D] = \frac{1}{2\pi} \left[ \int_D \text{Eu} \, dk_x \wedge dk_y - \int_{\partial D} A \right] \in \mathbb{Z} \tag{3}$$

upon also including a boundary term over $\partial D$. A nonzero value conveys that the braiding process has rendered band node charges that cannot be annihilated[8–10]. By creating band nodes and braiding them, the Euler class $|\epsilon|$ can moreover be transferred from one pair of bands to the next. However, in our Floquet system containing an infinite number of replicas, six bands are intrinsically divided into two similar branches per spectral period. The patch Euler class can transfer from the upper branch to the lower one as illustrated in Fig. 2b-2d, which is defined as Floquet Euler transfer (Fig. 2f), providing a window to novel invariants and physics beyond the equilibrium setting.

We now characterize the topological structure of non-Abelian frame charges in our scattering network. We map out in Fig. 2e the complete topological phase diagram for every possible realization of the scattering matrix $S_0$, represented by the angle parameters ξ and η. Its center corresponds to a singular phase, which is the perfect circulator case with $|T| = 1$ and infinite isolation. The yellow dotted lines represent phases preserving $\mathcal{T}$-symmetry, which shows a usual Kagome dispersion with quadratic nodes at the Γ point. Gaps 3,6 are opened gaps without any Dirac node so that the three-band branches are separate and well-defined. For static three-band systems, the charges take values in the quaternion group. In general, for $N$-band systems, however, the charges are governed by generalized quaternion charges that anti-commute between adjacent gaps. These charges are determined by the homotopy group of the flag manifold, which relates to the Salingaros vee group—conveniently expressed in terms of Clifford algebra basis elements[9,14,15,37]. In dynamical settings, this charge structure is naturally extended and captured by Dirac strings[14,39]. In our six band Floquet system we utilize $Cl_{0,6} = \{e_1, e_2, e_3, e_4, e_5, e_6\}$[39] and denote nodal charges as $\pm Q_{ij} = e_i e_j$, where $i$ and $j$ correspond to bands 1-6 and we identify $e_0$ with $e_6$, i.e. $Q_{12} = e_1 e_2$ and $Q_{61} = e_6 e_1$. Other conjugacy classes/charges are $\pm Q_{ijkl}$, and $\pm Q_{ijklmn}$. Physically, $\pm Q_{ij}$ represents $\pm \pi$ rotations in the plane spanned by the eigenvectors of bands $i$ and $j$. The frame charge $+1$ denotes a trivial rotation, and $-1$ refers to a $2\pi$ rotation in any rotation plane.

For the point-i in Fig. 2a, the quadratic node in gaps 4(1) have non-Abelian frame charge $-1$, whereas the Dirac nodes in gaps 5(2) have frame charges $\pm Q_{56}(Q_{23})$, respectively. When $\mathcal{T}$-breaking is implemented (e.g. point-ii in Fig. 2e), the quadratic node is split into four Dirac nodes with one frame charge $+Q_{45}(Q_{12})$ at the Γ point and three frame charges $-Q_{45}(Q_{12})$ nearby. These Dirac nodes can be created because they have a total frame charge $\mathbb{Q} = -1 = (-Q_{45})^3(+Q_{45})$ for gap 4 and $\mathbb{Q} = -1 = (-Q_{12})^3(+Q_{12})$ for gap 1 (Fig. 2b). Notably, these three $-Q_{45}(Q_{12})$ charged Dirac points not only move along the high-symmetry lines in the momentum space, but also construct a migrating Dirac energy contour (MDEC) that shifts along the Floquet quasi-energy dimension. Each MDEC moves down by one period of Floquet quasi-energy (by $2\pi$) with a circle around the singularity point of the phase diagram in Fig. 2e. Such a non-trivial migration is crucial for creating, braiding, merging, splitting and reconstructing the non-Abelian nodes, and thus our platform allows for richer Floquet topology phases to emerge (see Fig. 3 more details). The evolution of the non-Abelian frame charges of the band nodes in the parameter space $(k_x, k_y, LP)$ in Figs. 3b, 3c gives the phase diagram of the non-Abelian topological semimetal. It illustrates elegantly the topological transitions due to the migration of two MDECs.

At loop path (*LP*) point LP=3.0, we demonstrate that the Floquet system merges three bands from adjacent branches along the quasi-energy axis, giving rise to a phase in which all bands are connected. This defines an anomalous multigap phase that cannot exist in an equilibrium setting. Such a merging results in multi-gap triple degenerate nodes defined here as topological charge $-1$ at the quasi-energy of 0 and $\pi$ that have no static counterpart. After crossing this LP point, the lowest energy band of the upper branch is degressive to the highest energy band of the bottom branch. At the phase illustrated in Fig. 2c, six bands are all connected by two triple degenerate band touchings. The unconventional abrupt swap of the three moving Dirac nodes from $-Q_{45}(Q_{12})$ to $+Q_{34}(Q_{61})$ leads to the Floquet Euler transfer. The non-trivial Euler class is then transferred from gaps 4(1) to gaps 3(6), respectively. The manifestation of a non-trivial Euler class during a cyclic evolution in Fig. 2e is a stepwise reduction of the band gap, occurring as triply degenerate nodes appearing sequentially, as shown in Figs. 2f, 3c.

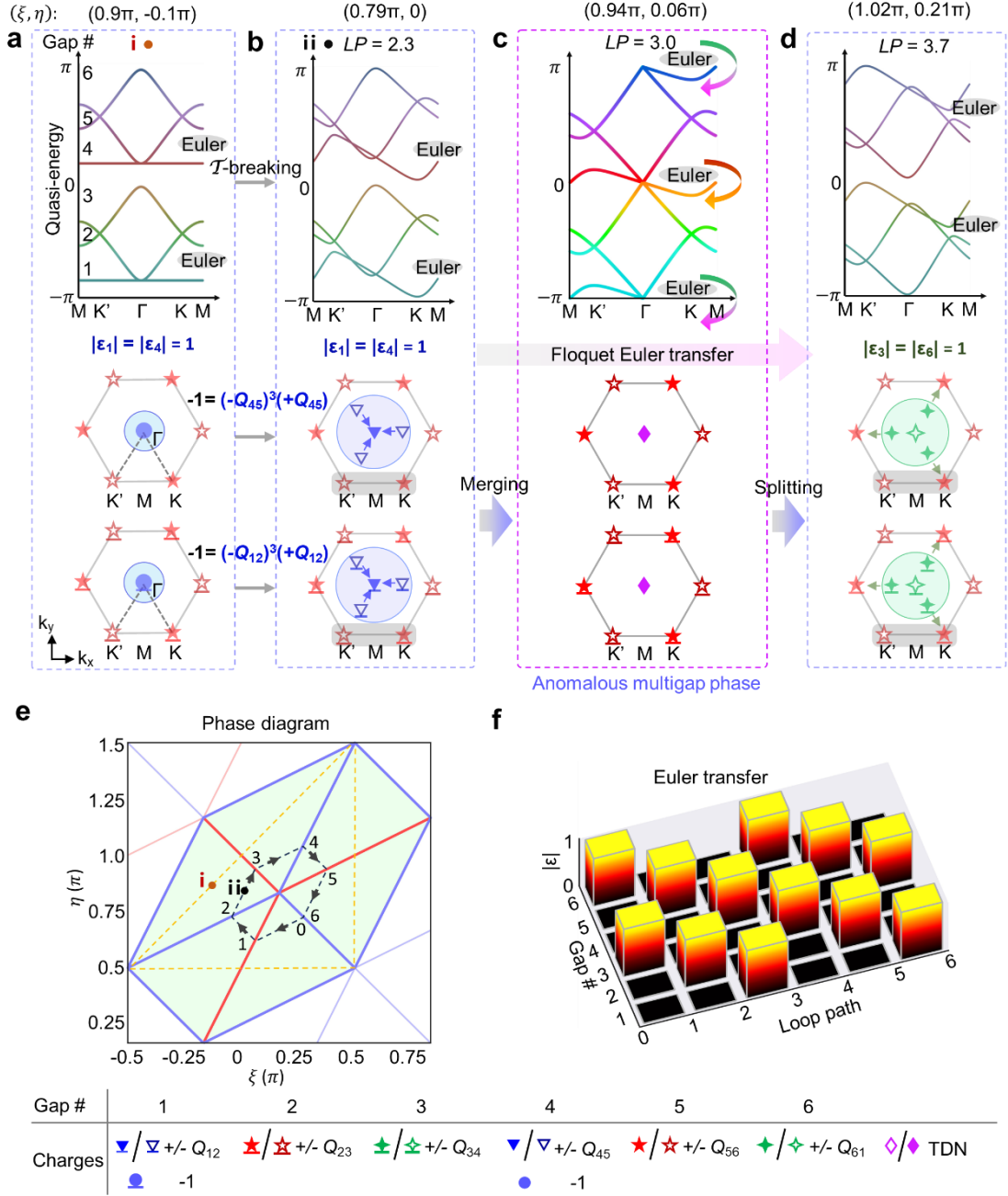

**Fig. 2 | Anomalous non-Abelian multigap phase and Floquet Euler transfer. a-b**, $\mathcal{T}$-breaking induced Floquet non-Abelian band topology and patch Euler class. Top: Quasi-energy band structures. Bottom: Non-Abelian node charges. Labels 1–6 indicate band gaps. Under $\mathcal{T}$-symmetry breaking (right), quadratic nodes split into one fixed and three mobile Dirac nodes, following quotient rule $-1 = (-Q_{45})^3(+Q_{45}) = (-Q_{12})^3(+Q_{12})$. **c**, Mechanism of Floquet Euler transfer induced by an anomalous non-Abelian multigap phase: the periodicity of quasi-energy cycles enables Dirac nodes existing at all band gaps at the critical point. As a result, the Euler class can be transferred, crossing the gap between adjacent branches. **d**, Multigap non-Abelian band topology after finishing the Floquet Euler transfer. The four nodes in gaps 1 and 4 (blue circle) are stably characterized by patch Euler class $|\epsilon_{1,4}| = 1$, which is then transferred into gaps 6 and 3 (green

circle) with $|\epsilon_{6,3}| = 1$ crossing the Floquet spectral boundaries. **e**, Topological phase diagram in (ξ, η) space. Green-shaded region: minimal periodic zone in the parameter space. Green dotted/blue/red lines: $\mathcal{T}$-symmetric phases/conventional Euler class transfer/Floquet Euler transfer. Dark dotted loop: path encircling a band singularity. **f**, Euler class evolution within each band gap as a function of the loop path parameter.

Using the outlined strategy, we now analyze the Euler class transfer, Dirac nodes, equal energy contour evolution that is induced by circling around the singularity in the phase diagram. The iso-energy contour of Dirac nodes within one Floquet cycle provides a particularly elegant indicator to visualize the non-trivial role of MDECs that are split from the quadratic points (Fig. 2b). In our photonic scattering network, breaking $\mathcal{T}$-symmetry results in four distinct Dirac iso-energy contours. Among these, Dirac nodes are located at the Γ, K, and K' points of the Brillouin zone. Additionally, there exists a degenerate iso-energy contour shared by three displaced Dirac nodes, marked as MDEC1 and MDEC2 for the two branches. As shown in Fig. 3a, the iso-energy contours for the momentum space fixed Dirac nodes (at Γ, K, and K') do not show observable topological migration features along the quasi-energy dimension. After completing a loop encircling the singular point in the parameter space, the iso-energy contours return to the initial level. However, as for the MDEC1 and MDEC2, both iso-energy contours undergo a full-period shift in Floquet quasi-energy spectrum, akin to a topologically nontrivial 'pumping' process of the migrating Dirac cones (Fig. 3a and Supplementary Information Fig. S2).

With a pivotal role played by adjacent gaps in multi-gap topology[8–10] it is compelling to address the effect of topological transitions across adjacent gaps. We monitor the Euler phase summarized in Fig. 2f, where the patch Euler class is transferred between different subspaces (band nodes) by involving *all gaps* in the spectrum. We find that the non-trivial Euler class is closely related with the moving Dirac cones. Since the gaps are cyclic in the quasi-energy dimension, the Euler class descends through band gaps one by one and finish one cycle with a complete loop path. We further demonstrate the detailed energy band dispersions along the loop path, which clearly visualize the relationship among multiple bands, gaps, Dirac nodes and MDECs (Fig. 3c and Extended Data Figs. 4-5). One can see that their main characteristic is that the topological phase transitions of all Dirac nodes are highly correlated to the two MDECs.

Taking the $LP = 3.0$ in Fig. 2c as an example, triply degenerate nodes connecting all Floquet bands occurs at the Γ point. After crossing this phase, the lowest band of each branch ($LP < 3.0$) is transferred onto the adjacent branch ($LP > 3.0$), where the new branches are now separated by a gap.

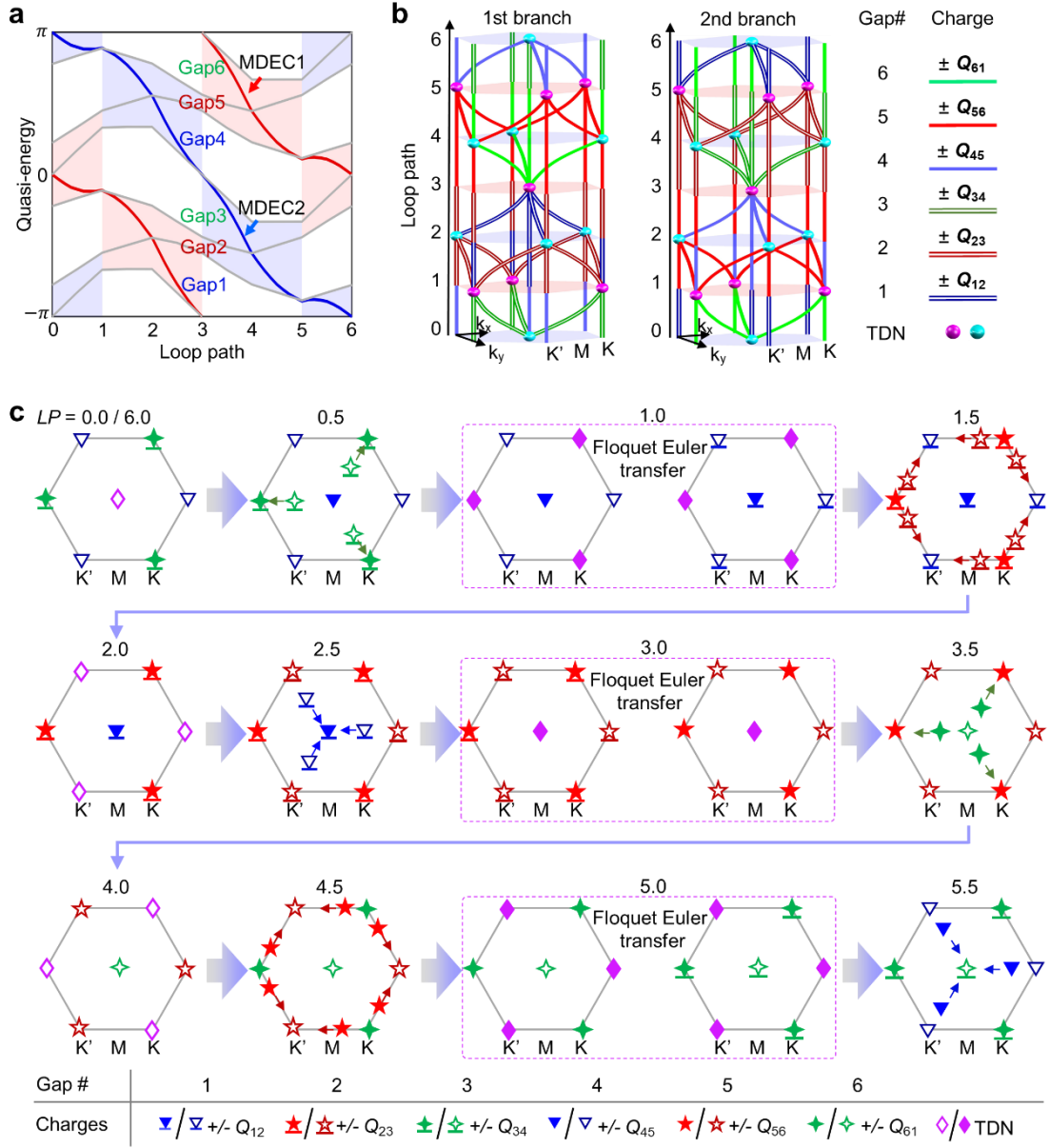

**Fig. 3 | Evolution of quaternion charges. a**, Evolution of quasi-energy contours for all Dirac nodes. Red and blue curves correspond to the first migrating Dirac energy contour (MDEC1 in the 1st branch) and to the second one (MDEC2 in the 2nd branch), respectively. The MDEC indicates the quasi-energy contour that harbors three moving Dirac nodes, showing a full-spectrum pump within one loop path cycle due to the Floquet periodicity. **b**, Non-Abelian charge evolution versus loop path for the two branches, showing transitions across multiple topological phases. **c**, Evolution of the non-Abelian charges for the 1st branch by cycling one round along the loop path in Fig. 2e.

**Anomalous Dirac strings and Floquet-induced braiding**

We further showcase that novel anomalous topological phases may arise by an interplay of multi-gap topological principles and the periodicity of the Floquet Brillouin zone in the extended Data. Extended Data Fig. 1 displays examples of new anomalous Dirac string phases, which are in fact ubiquitous in our photonic scattering network. The Floquet spectrum of our Kagome-based scattering network exhibits a distinctive structure consisting of two branches. Each branch consists of three bands interconnected by Dirac points, and the two branches are separated from each other by band gaps [see Fig.1d]. Interestingly, when Dirac strings traverse these band gaps, they are naturally partitioned, resulting in the formation of two anomalous Dirac strings, as shown in Extended Data Fig. 1a. This architecture thus provides a native platform for studying exotic topological entities.

Concretely, the Dirac string configurations relate to the Zak phases of each band. A Dirac string threading through a band gap imposes a universal signature: a π-phase shift in the Berry phase of the adjacent bands. Distinct from their static counterparts, Floquet band structures exhibit a periodic interplay between the Zak phases of the bands and the Dirac strings in the gaps[14]. This is exemplified by the global Dirac strings spanning gap 3 and gap 6, which play a critical role in locking the Zak phase distributions of the upper and lower sets of bands into alignment, as shown in Extended Data Figs. 1e-1h. We note the discrepancies in the local Dirac string configurations within the respective gaps of each branch, underscoring a different topological signature between two branches, signaling a faithful six-band system. Along the path loop in parameter space, anomalous Dirac strings transfer between different band gaps. This process primarily stems from the Floquet Euler transfer mechanism, during which the two spectral branches undergo topological reconstruction, leading to the opening of gaps at distinct quasi-energy windows. The detailed evolution can be elucidated by analyzing the anomalous multigap phase and the corresponding Dirac-string configurations, as illustrated in Extended Data Figs. 1b–1c and 1f–1g. Specifically, anomalous Dirac strings initially residing in gap 6 and gap 3 migrate into the upper band of the first and second branches, respectively. Meanwhile, those originally located in gap 4 and gap 1 migrate into the band gaps, thereby manifesting as robust anomalous Dirac strings, as shown in Extended Data Figs. 1d, 1h.

The braiding of band nodes represents a key topological feature of two-dimensional non-Abelian systems and is closely linked to Euler transfer. We next examine the interplay between Euler class transfer, band node braiding, and their connection to the anomalous Dirac strings that arise uniquely in Floquet settings. As a representative case for the first branch (Extended Data Fig. 2a), at $LP = 2.5$, gap 1 exhibits a nontrivial Euler class, characterized by frame charges decomposing into two distinct pairs: $(-Q_{12})(+Q_{12})$ and $(-Q_{12})(-Q_{12})$. The evolutionary paths of these pairs diverge: the $(-Q_{12})(+Q_{12})$ pair undergoes direct annihilation at the $\Gamma$-point, while the $(-Q_{12})(-Q_{12})$ pair combines there with a band from the adjacent branch, collectively forming a stable triple degeneracy (Extended Data Fig. 2b). This triple degeneracy subsequently splits asymmetrically in gap 6, generating a topologically trivial $(-Q_{61})(+Q_{61})$ pair and a newly created $(+Q_{61})(+Q_{61})$ pair, with the latter pair carrying a nontrivial Euler invariant (Extended Data Fig. 2d). A key intermediate step involves the mobile $+Q_{61}$-charged Dirac points in Gap 6 propagating from $\Gamma$ toward K. Crucially, two of them cross the anomalous Floquet Dirac strings residing in gap 1—a direct consequence of braiding from the periodic Floquet drive—which induces a deterministic reversal of their topological charges (Extended Data Fig. 2e). This prepared state then enables a subsequent recombination event at the K-point, where these mobile Dirac points merge with a resident $+Q_{56}$-charged Dirac node in Gap 5. This recombination triggers a further cycle of splitting and pair creation (Extended Data Figs. 2f, 2g). Finally, the resulting mobile $-Q_{56}$ Dirac point in gap 5 traverses a Dirac string within gap 4, experiencing a final sign reversal of the frame charge that collectively renders the Euler class in gap 5 nontrivial. Thus, through a precisely orchestrated sequence of recombination, annihilation, creation, and braiding of Dirac points—mediated by crossings with anomalous Floquet Dirac strings—the nontrivial Euler class is effectively transferred from gap 1, through gap 6, and ultimately resides within gap 5. Similarly, the second branch exhibits a parallel evolution as illustrated in Fig.2f, and further detailed in Extended Data Fig. 3.

**Correspondence between Floquet edge states and multigap Dirac strings**

Dirac strings serve as robust predictors of anomalous edge states by explicitly marking the gauge discontinuities arising from band inversions during nodal phase transitions[14], as shown in Fig. 4. Their configuration directly determines the π Zak phase accumulation along paths perpendicular to the edge. This quantized shift signals a topologically non-trivial gap and guaranteesthe existence of localized edge states within that gap (see Extended Data Fig. 6 for the detailed correspondence between Dirac strings and edge states). Therefore, tracking Dirac strings provides a universal and systematic approach to predicting boundary states across different gaps in the spectrum.

As pointed out in [14], we can effectively check for edge states using the Zak phases along contractible paths in the Brillouin zone. The extra Dirac strings provide phase jumps and when such a jump is π (mod 2π) edge states emerge (see Supplementary Figs. S3-S5). Importantly, by effectively keeping track of the bulk-edge correspondence in the above-described rules (see Fig. 3 and Extended Data Figs. 4-5 for the whole evolution process), our analysis further reveals a defining characteristic of the unconventional Floquet multiband antichiral edge states within a band. We stress that with the nomenclature antichiral we mean that the dispersion is opposite with respect to momentum; and not chiral edge states transversing gaps that give a finite quantized transport which do not appear in multi-gap topological antichiral modes[40,41]. In Fig. 4, we present comprehensive edge-projected bands and their corresponding edge-state patterns for the evolution of the whole loop path. As revealed by our previous work[7], the multigap antichiral edge states can be tuned versatilely with different $\mathcal{T}$-symmetry breaking strength for a static Kagome lattice. For six bands with four connected by Dirac nodes and two opened ones in this network, the Dirac string configuration is determined and then the (dis-)appearance of corresponding edge states in each gap has been successfully characterized. At the critical point when Euler transfer occurs (Figs. 4b, 4c), Dirac string patterns in gaps 1, 3, 4 and 6 can be all considered as anomalous as all gaps in the Floquet spectrum are involved topologically. The anomalous Dirac strings in gap 1 and gap 3 can be projected to the edge along *x*-direction and thus lead to the coexistence of multigap edge states in three adjacent bands from 1 to 3. However, the Dirac strings in gap 4 and gap 5 are perpendicular to the *x*-directed edge, showing no edge correspondence. Without the unique feature of anomalous Floquet phases, the edge state distributions cannot be predicted anymore if we simply focus on an isolate branch of bands. A direct mapping of the boundary modes predicted by the theory over the complete loop is provided in Extended Data Fig. 6.

In our scattering network, the periodicity of Floquet spectrum and evolution of two MDECs induce some unconventional edge states. As shown in Extended Data Fig. 7, a pronounced signature is that edge states in two branches of three bands exhibit in complementary gaps with opposite group velocities. This arises because: the y-direction boundaries possess fixed termination phases; concurrently, the two branches exhibit a π quasi-energy difference, thus inducing antichiral edge states in a band gap to emerge on opposing sides of the system terminated by Dirac points. Another hallmark signature is that the MDECs always act as a start- or end-point that create or terminate one antichiral edge state, respectively. This behavior is traced back to the projection of the three Dirac nodes in the MDEC onto the $k_x$-axis which results in two nodes being degenerate in momentum space, leaving a single, isolated Dirac node at a distinct $k_x$ value. This isolated node induces a switch in the Zak phase between 0 and π. When the MDEC is in the bottom gap of each branch, its decrement would shorten the length of antichiral edge states, such as in the loop path ranges 0~1, 2~3, 4~5. After crossing the points when $LP$ = 1,3,5, the MDEC is transferred into the upper gap, its decrement would increase the length of antichiral edge states, as shown in the loop path ranges 1~2, 3~4, 5~6. We note that when cycling one loop path, the edge states are also 'pumped' down one cycle of Floquet spectrum, as shown in Extended Data Fig. 7, highlighting the rich topological interplay in our system.

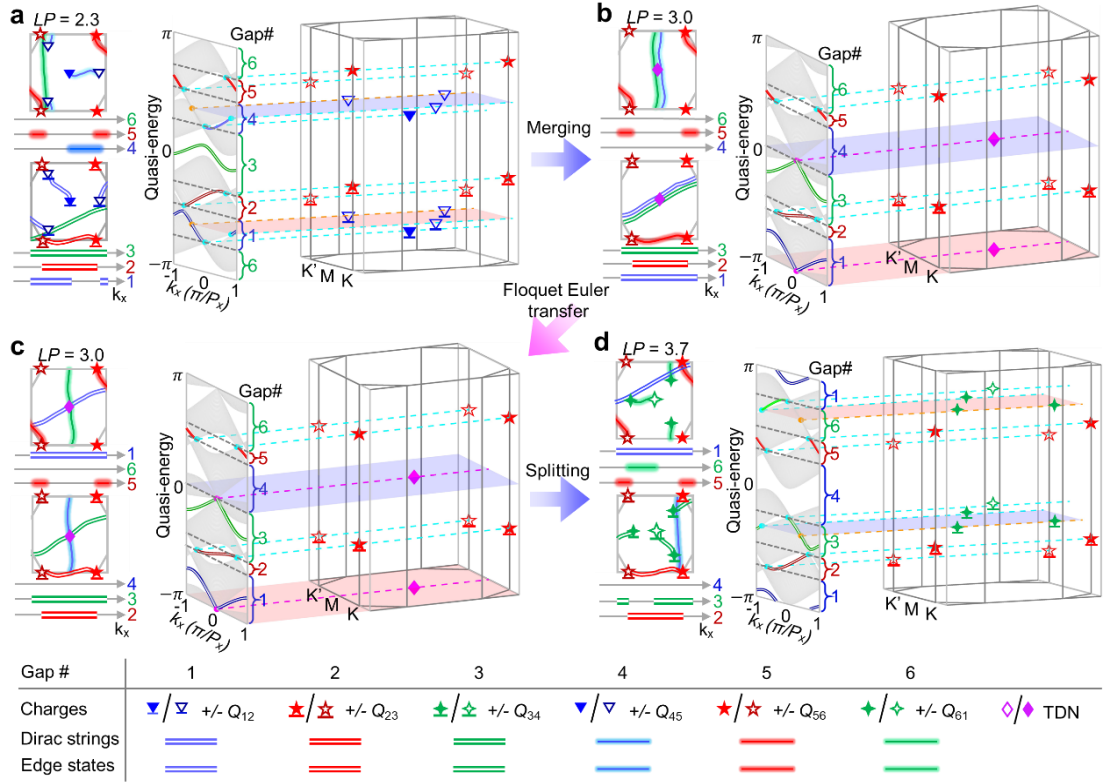

**Fig. 4 | Correspondence of multi-gap bulk quaternion charges, Dirac strings and topological edge states. a-d**, Bulk non-Abelian nodes, Dirac strings, projected quasi-energy bands and Brillouin zone for $LP = 2.3$ (a), $LP = 3.0$ (b, c) and $LP = 3.7$ (d). Left panels are the topological node arrangements and Dirac strings in all gaps. Line segments aligned along the $k_x$-axis mark the regions that possess edge states in each gap. Right panels show the projected edges and bulk non-Abelian charges.

**Experimental realization of 2D Floquet non-Abelian band topology**

To experimentally validate our theoretical framework, we constructed a photonic scattering network in a ribbon geometry (Fig. 5c), in which both the frequency dispersion of the scatterers and delay lines should be taken into account. The measured CW&CCW transmittances and reflection spectra of an individual circulator (Fig. 5a) are demonstrated in Fig. 5b, which matches well with the theoretical prediction at $LP = 3.0$. The spectra show a low dispersion in a broadband frequency range from 4.9 - 7.2GHz, thus ensuring the feasibility of Floquet spectrum measurements with more than one cycles. These kinds of spectra also appear in the phases with the $LP = 3.0$ shown in Fig. 5e. Fig. 5d illustrates the detailed configuration within one unit cell, two circulators with opposite magnetic flux are connected by coaxial cables in a Kagome graph. The scatterers are ferrite circulators connected with coaxial cables as shown in

Fig. 5f. The length of each coaxial transmission line determines the phase delay of the links changing linearly along the frequency dimension, as shown in Fig. 5g. The lines at the y-direction edges are cut to induce an on-site mirror-like reflection due to a strong impedance mismatch with the air.

When a source is located at the middle along *x*-direction and near at the two edges in y-direction, the excited edge states propagate along $\pm x$-directions. Via the Fourier transformation of the detected wavefunctions at each excitation frequency, we obtain the dispersions of the electromagnetic Bloch waves. Figure 5i shows that the measured electromagnetic band structures agree excellently with the theoretical predictions where two cycles of Floquet spectrum are provided in the Fig. 5h when $LP = 3.0$. We observe over a broad range of experimental parameters that the measured spectra match well with the theoretical band structures. In the experimental results, the dispersion of both the anomalous edge states and Floquet antichiral edge states are clearly visible. Particularly, let us consider the transport properties of field maps at the four discrete frequencies where antichiral edge states appears. As shown in Fig. 5j, the edge modes excited at both upper and bottom boundaries at the same frequency exhibits a typical antichiral feature and propagate along the same direction. The edge transport properties exhibit a striking periodicity between the first two and last two frequency points, despite opposite transport directions for adjacent ones. This manifests as distinct two-period Floquet characteristics. All these experimental observations fully confirm the theoretical findings as shown in Fig. 5. We note that this also presents the first demonstration for the periodic edge states associated in Floquet non-Abelian topological phases beyond static settings.

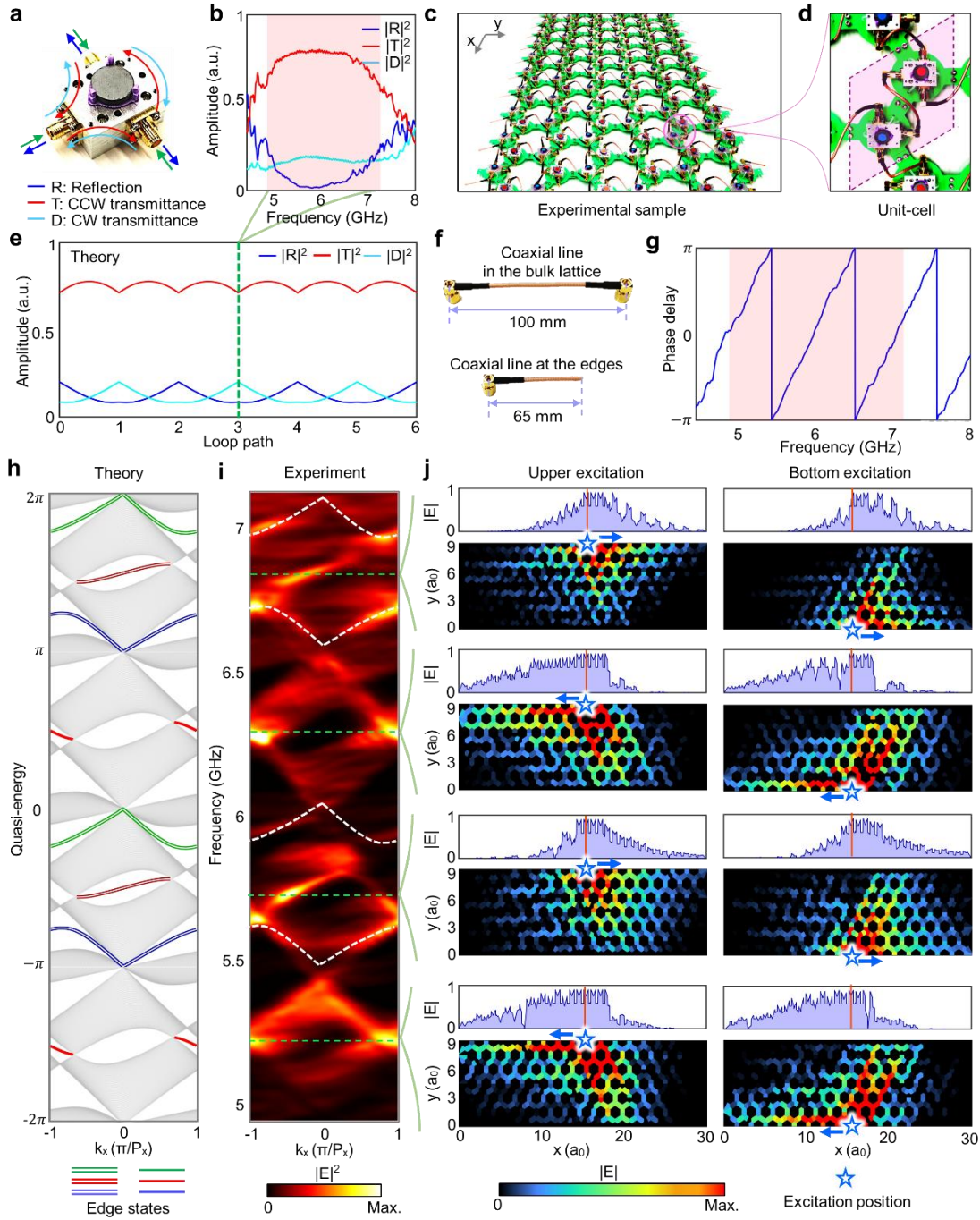

**Fig. 5 | Experimental realization of 2D Floquet non-Abelian band topology. a**, Photograph of a ferrite circulator implementing non-reciprocal scattering. **b**, Measured scattering spectrum of an individual circulator, showing broadband uniformity (4.9-7.2GHz). Parameters correspond to point $LP = 3.0$ in the $(\xi, \eta)$ phase diagram (Fig. 2a). **c**, Schematic of the experimental non-Abelian photonic scattering network. **d**, Unit cell structure comprising two circulators with opposing magnetic flux orientations, preserving $C_{2z}\mathcal{T}$ symmetry while breaking $\mathcal{T}$-symmetry. **e**, Theoretical predictions of one scatter for CCW transmittance (T), CW transmittance (D) and reflection (R) spectra along the loop path in Fig. 2a. **f**, Upper panel: coaxial transmission lines in the bulk lattice

acting as links to impart frequency dependent phase delays between adjacent circulators. Bottom panel: the coaxial transmission lines at y-direction edges are cut off to introduce boundary conditions with mirror reflection. **g**, Measured phase delay spectrum of two adjacent nodes. The red-shaded area represents the bandwidth of around $4\pi$ phase covering that corresponds to two cycles of Floquet quasi-energy. **h**, Theoretical projected band structure at $LP = 3.0$. Green and black lines denote Floquet antichiral edge states and anomalous edges, respectively. Two periodic Floquet zones are shown for the experimental verification. **i**, Experimentally measured band structure with excitations at upper and bottom edges (along y-direction). The white dotted lines highlight the Floquet anomalous edge states that reside between the two branches. Periodic antichiral edge states with staggered propagation direction are observed between 4.9-7.2GHz, matching theoretical predictions in (e). Green dashed lines mark four frequencies exhibiting antichiral edge states, highlighting Floquet periodicity. **j**, Experimentally measured electric field distributions (x-y plane) and edge field profiles (along x-direction) at the antichiral edge state frequencies. Left panels: excitation at y-direction upper edge. Right panels: excitation at y-direction bottom edge.

**Conclusion and Outlook**

In summary, our work demonstrates that photonic scattering networks provide a practical and experimentally tunable platform for realizing anomalous 2D Floquet non-Abelian topology. Using this architecture, we directly observe anomalous edge states spanning multiple quasi-energy gaps, an unambiguous experimental signature of multi-gap topology in a driven system. The underlying six-band geometry supports an anomalous multi-gap phase in which Dirac nodes link all Floquet bands and mediate the transfer of Euler invariants, as well as gapped Dirac-string phases; in both cases, the periodic drive compels the band nodes to get intertwined in a non-Abelian manner, resulting in a redistribution of topological charge throughout the spectrum. In additional, this driving-enabled Euler transfer induces a Floquet-periodic pumping effect, highlighting unique bulk signatures. These results establish a coherent connection between Floquet dynamics, non-Abelian band topology, and experimentally measurable boundary modes, revealing how periodic driving can activate topological effects inaccessible in static systems. More broadly, our work showcases the promise of driven multi-gap platforms for engineering out-of-equilibrium phases and enabling new approaches to wave control across photonic, acoustic, electronic, and quantum systems.

**Methods**

**Non-Abelian charge and Dirac string calculation**

We compute eigenstate trajectories across the loop path-tuned synthetic space ($k_x, k_y, LP$) — formed by augmenting the 2D Brillouin zone with the loop path dimension — to unveil nodal quaternion-valued frame charges (Fig. 2c). Each nodal encirclement follows an anticlockwise loop with fixed basepoint $A_0$. Path construction initiates near the node with flux incrementation establishing the trajectory origin, followed by adiabatic circulation. Projected trajectories for the eigenstates of three bands that connected by nodes then reveal topologically protected phase windings, and then the non-Abelian charge can be determined.

To determine the spatial distribution of Dirac strings within the anomalous Dirac string phase, we computed the Zak phase for each band along the path perpendicular to three zigzag edge directions in the Brillouin zone. The Zak phase is evaluated by integrating the Berry connection over closed, non-contractible paths aligned with the reciprocal lattice vectors[3]. The resulting pattern of the Zak phase reveals the locations of the Dirac strings since a π-phase shift in the Zak phase along a given path indicates that the integration contour crosses a Dirac string. By analyzing the Zak phase distributions of six bands, we can identify Dirac strings between respective band pairs, with two anomalous Dirac strings specifically spanning the gaps between the two branches, as summarized in Supplementary Information Figs. S3-S5.

**Experimental design and characterization**

Microwave coaxial lines are employed as phase-delay links throughout the network, with precisely engineered lengths to achieve targeted topological phases while maintaining a $50\,\Omega$ characteristic impedance for optimal power transfer. The phase delay $\varphi$ introduced by each coaxial segment of physical length $L$ at operational frequency $f$ via propagation delay measurements — accounting for the frequency dispersion. To navigate practical constraints imposed by circulator dispersion, we constructed an experimental map of the theoretical model at $LP = 3.0$ (see Supplementary Information Fig. S6) parameterized as $L = 100$ mm and $f$ scanning from 4.9 GHz to 7.2 GHz to robustly access Floquet edge states for two cycles of quasi-energy spectrum.

Scattering parameters and field distributions across all fabricated networks are characterized using a vector network analyzer (VNA) with precision coaxial connections. For multiport scattering measurements, two network ports were connected to the VNA under through-reflect-line (TRL) calibration, while remaining ports terminated in matched 50-Ω loads to suppress reflections. Edge transport properties are quantified by putting the excitation at the boundaries in the y-direction. Field mapping is performed by fixing a signal input port to VNA channel 1 and manually scanning field amplitudes and phases at critical nodes using a calibrated coaxial probe connected to channel 2. Then, the projected bulk and edge band dispersions are reconstructed through two-dimensional Fourier analysis of the spatially resolved complex field measurements at discrete frequency points. This integrated approach enabled simultaneous acquisition of band structures, edge transport efficiencies, and topological signatures within a single experimental framework.

## Acknowledgments

The work was sponsored by the Key Research and Development Program of the Ministry of Science and Technology grant 2022YFA1405200 (Y.Y.), 2022YFA1404900 (Y.Y.), 2022YFA1404704 (H.C.), and 2022YFA1404902 (H.C.), National Natural Science Foundation of China (NNSFC) grant 62175215 (Y.Y.), 61975176 (H.C.), 62305384 (Y.H.), 62305298 (M.T.), China National Postdoctoral Program for Innovative Talents grant BX20230310 (M.T.) Youth Innovation Talent Incubation Foundation of National University of Defense Technology grant 2023-lxy-fhij-007 (Y.H.) Key Research and Development Program of Zhejiang Province grant 2022C01036 (H.C.), and the Class D for Young Scientific Research Peak Creation No.K20250230 (Y.Y.). R.-J.S. acknowledges funding from an EPSRC ERC underwrite grant EP/X025829/1, and a Royal Society exchange grant IES/R1/221060. F.N.\"U. acknowledges support from the Royal Society Grant~URF/R1/241667 and Trinity College Cambridge. The work of R.-J.S. and F.N.\"U. was performed in part at Aspen Center for Physics, which is supported by National Science Foundation grant PHY-2210452.

## Author contributions

Y.H., Y.Y., R.-J.S., F.N.Ü., and M.T. created the design; Y.H., Y.Y. and M.T. designed the experiment and fabricated samples; Y.H. measured the data with assistance from M.T., S.Y., N.H., F.C., L.Z., R.Z., and Q.C.; R.-J.S., F.N.Ü., and Y.H. provided the theoretical explanations; Y.Y., R.-J.S., F.N.Ü., T. J., and H.C. supervised the project; Y.H., and M.T. wrote the manuscript with input from Y.Y., R.-J.S., F.N.Ü., and H.C.

## Competing interests

The authors declare no competing interests.

**Data and materials availability**

All data that support the findings of this study are present in the paper and the Supplementary Information. Any additional information may be available from the corresponding authors upon request.

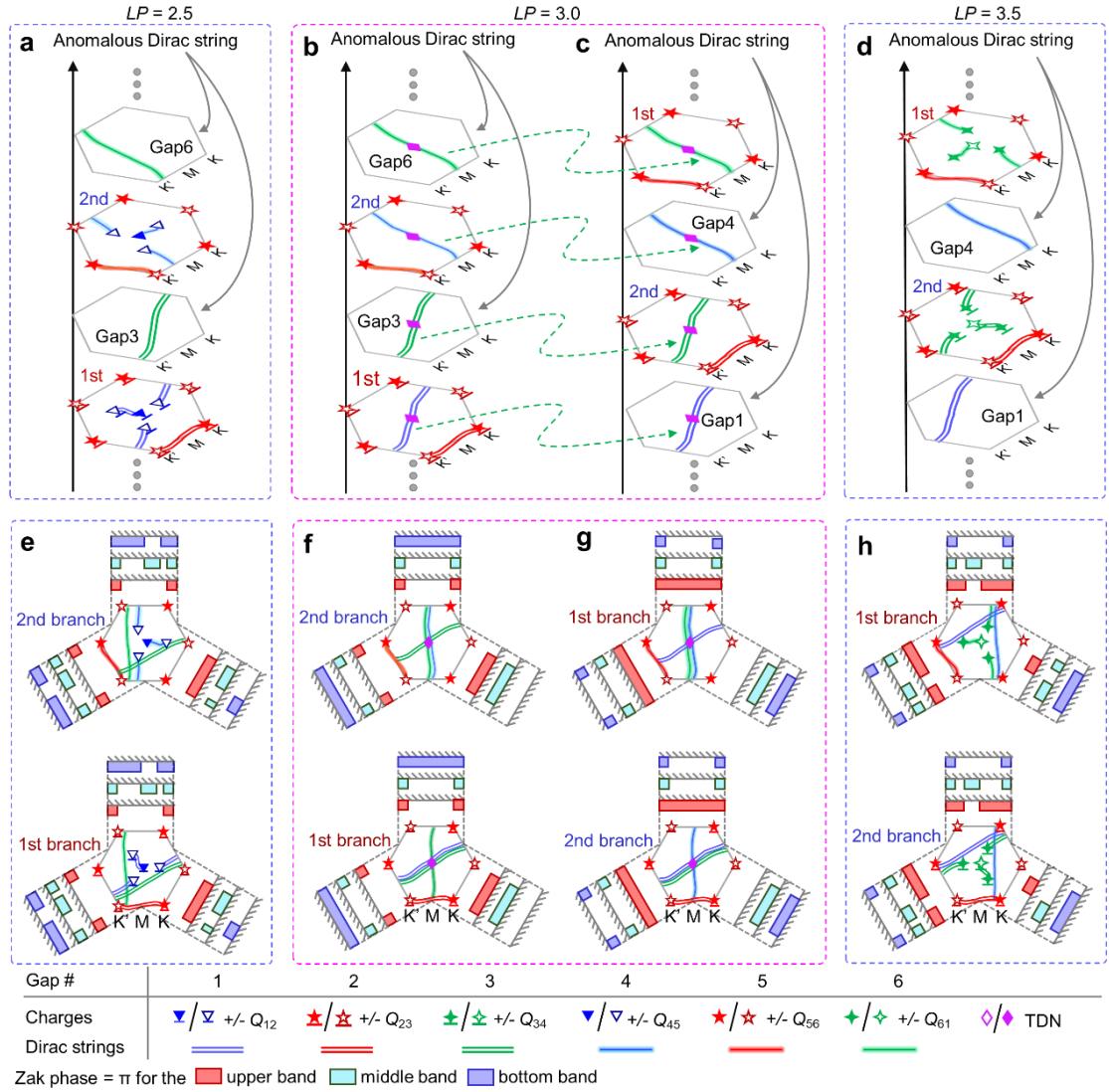

**Extended Data Fig. 1 | Anomalous Dirac strings and Zak phase distributions. a-d**, Evolution of the bulk non-Abelian charges and Dirac stings in the loop path ranging from 2.5 to 3.5. Here, the 1st represents the first branch in which the MDEC1 exists, and the 2nd corresponds to the second branch in which the MDEC2 exists. **e-h**, Detailed connections between Dirac strings and projected Zak phases along the loop path from 2.5 to 3.5. The topology of the three zigzag edges is quantified by the non-trivial values of the Zak phase, which is integrated along high-symmetry paths traversing the Brillouin zone in the direction normal to the edges.

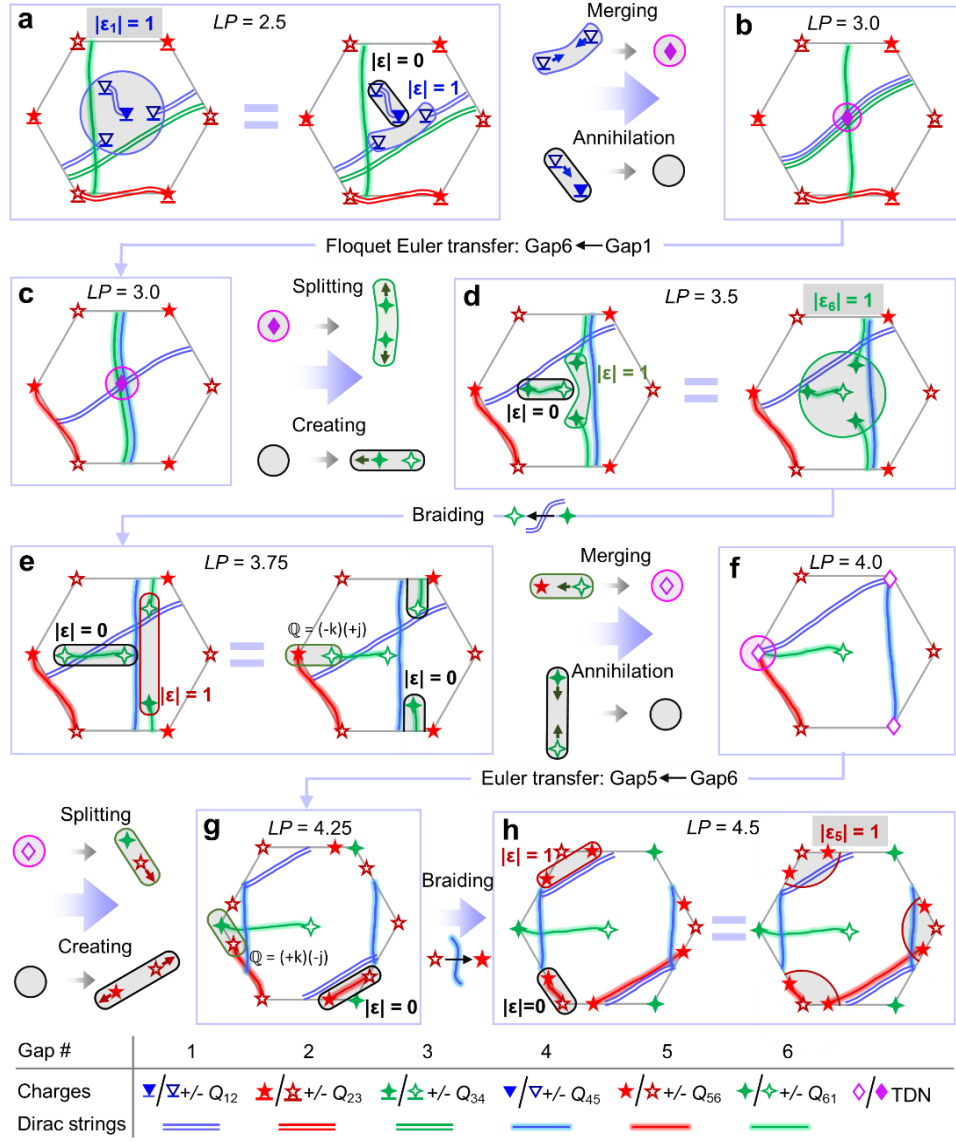

**Extended Data Fig. 2 | Braiding of band nodes in adjacent gaps and Euler transfer for the 1st branch.** The resulting braiding configuration is visualized through the graphical depiction of non-Abelian frame charges and their associated Dirac strings. The stability of nodal points during recombination (braiding) within designated patches is characterized by the patch Euler class. The large circles and ellipses outlined represent the patches used for computing the patch Euler class.

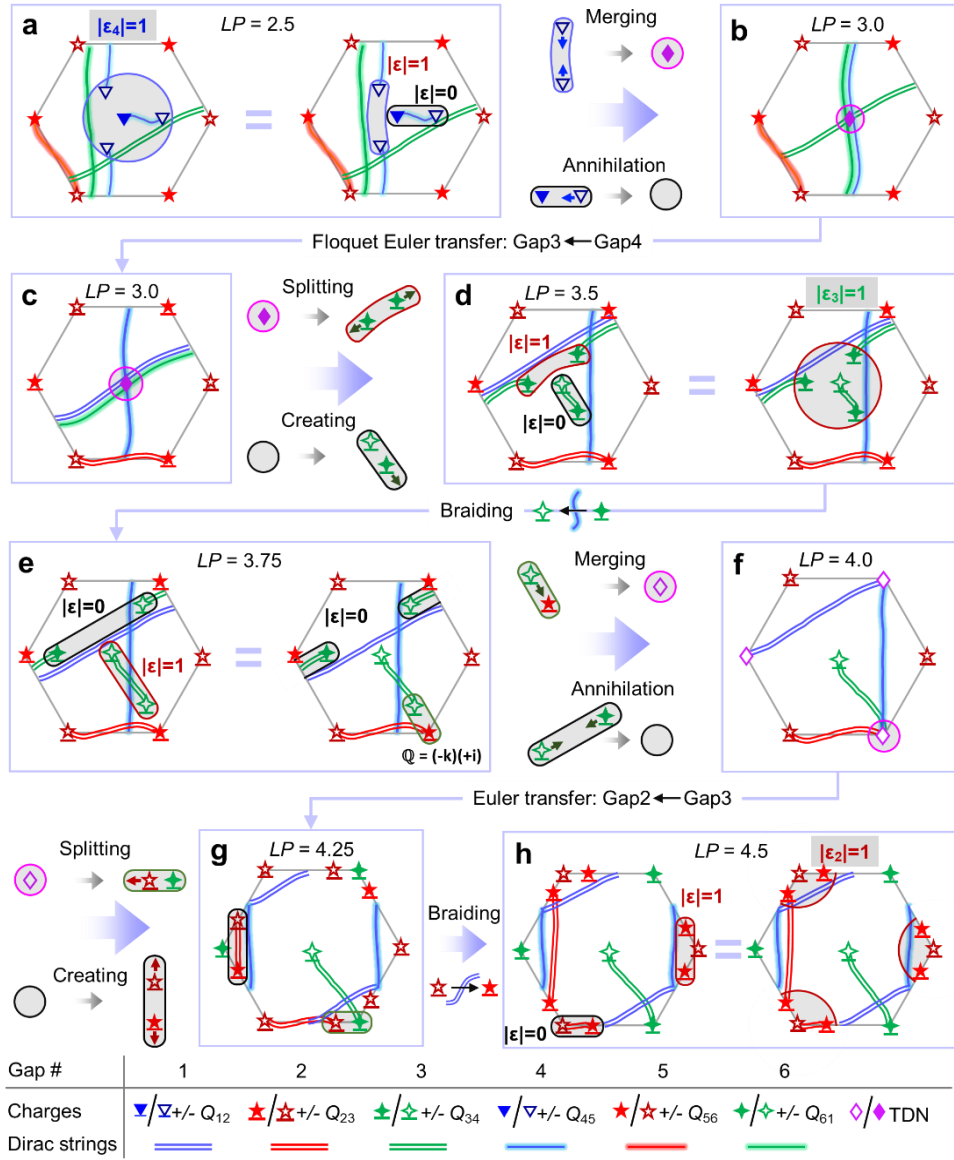

**Extended Data Fig. 3 | Braiding of adjacent nodes in gaps and Euler transfer for the 2nd branch.** The resulting braiding configuration is visualized through the graphical depiction of non-Abelian frame charges and their associated Dirac strings. The stability of nodal points during recombination (braiding) within designated patches is characterized by the patch Euler class. The large circles and ellipses outlined represent the patches used for computing the patch Euler class.

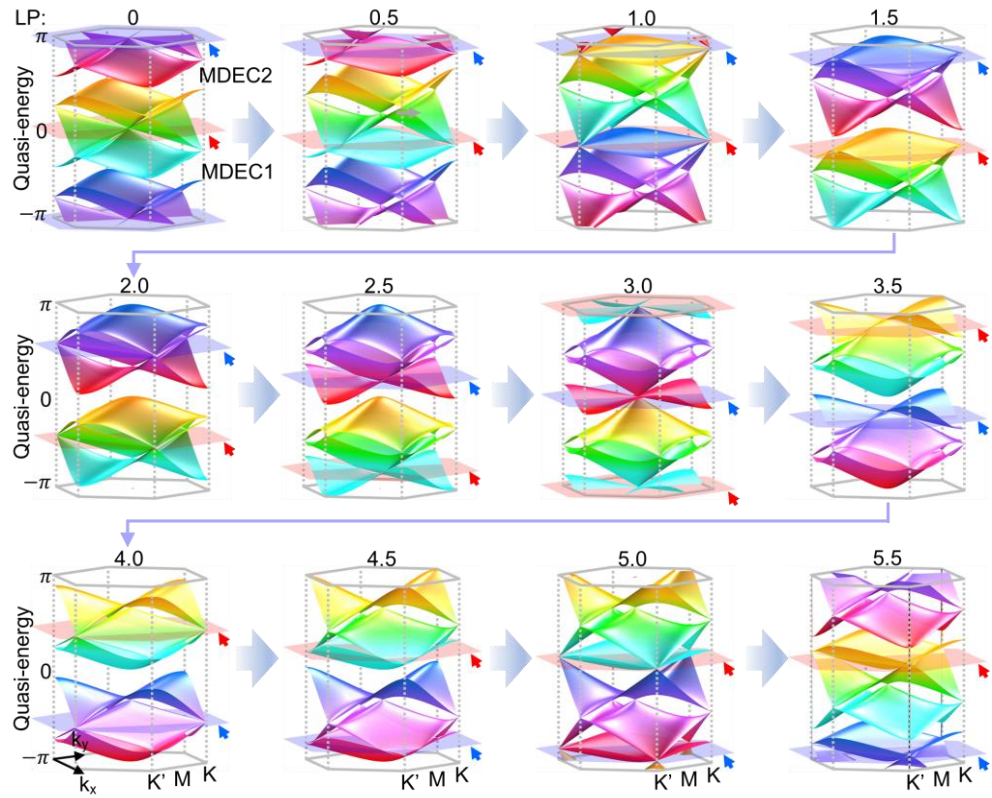

**Extended Data Fig. 4 | Evolution of the bulk band dispersions over one cycle along the loop path in Fig. 2e.**

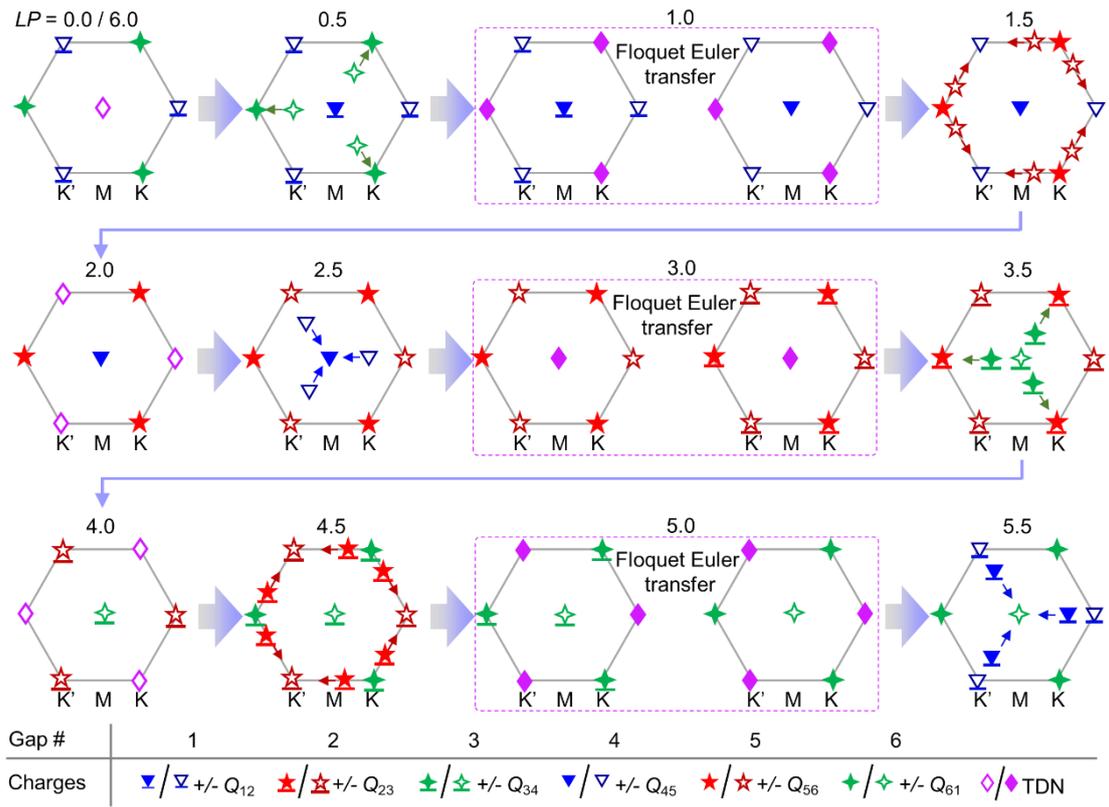

**Extended Data Fig. 5 | Evolution of the non-Abelian charges for the 2nd branch over one cycle along the loop path in Fig. 2e.**

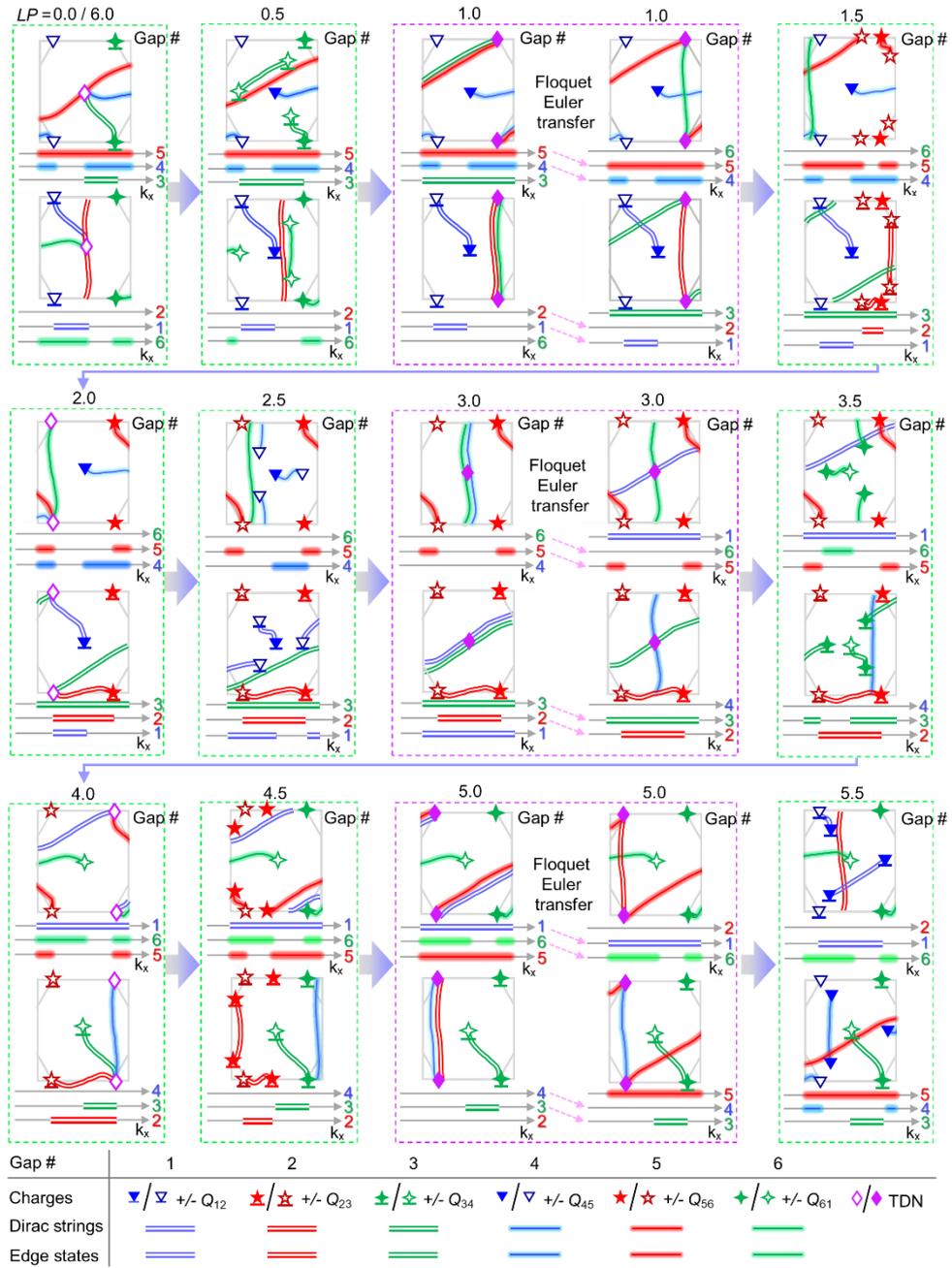

**Extended Data Fig. 6 | The correspondence between edge states and Dirac strings along the loop path in Fig. 2e.** Line segments aligned along the $k_x$-axis mark the regions that possess edge states in the gaps, where the existence of Dirac strings leads to the edge states through projections perpendicular to the edge direction.

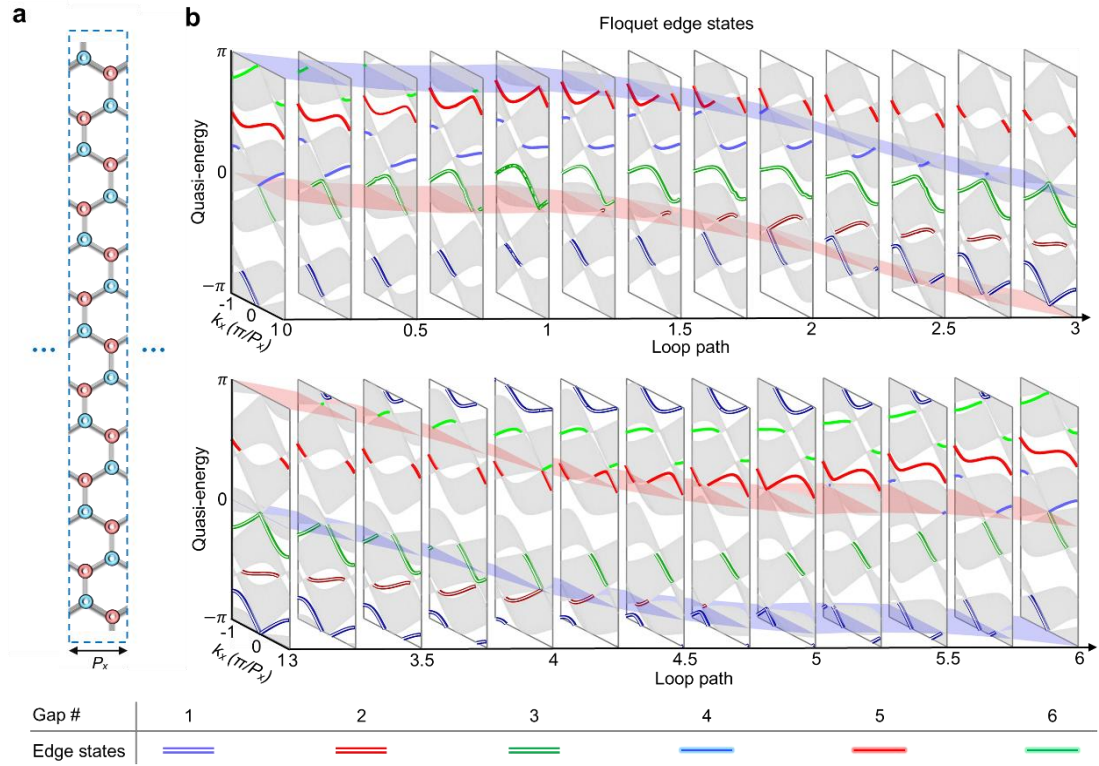

**Extended Data Fig. 7 | Evolution of the Floquet edge states. a**, In the real-space ribbon geometry considered, the unit cell is periodic along the *x*-direction and terminated along the y-direction, consequently resulting in the formation of one-dimensional edge states localized at the parallel boundaries along *x*. **b**, Band structure evolution of a supercell illustrated in (a).

Supplementary Information for

# Realizing anomalous Floquet non-Abelian band topology in photonic scattering networks

**I. Theoretical Nonreciprocal Scattering Modelling for C₃-Symmetric Circulators**

The core scattering element comprises a lossless resonant cavity with threefold rotational symmetry ($C_3$), subjected to an external time-reversal symmetry breaking bias (e.g., magnetic field) that lifts degeneracy via Zeeman-like splitting. This splits the fundamental mode into right-handed ($\omega^+$) and left-handed ($\omega^-$) eigenfrequencies, establishing the foundation for nonreciprocity.

Temporal coupled-mode theory yields a unitary scattering matrix $S_0 \in 3 \times 3$ governing wave dynamics:

$$\boldsymbol{S_0} = \begin{bmatrix} R & T & D \\ D & R & T \\ T & D & R \end{bmatrix} \quad (S1)$$

where $R = -1 + \frac{2}{3}(cos\xi\, e^{i\xi} + cos\eta\, e^{i\eta})$, $T = \frac{2}{3}\left(e^{-\frac{i2\pi}{3}}cos\xi\, e^{i\xi} + e^{\frac{i2\pi}{3}}cos\eta\, e^{i\eta}\right)$, $D = \frac{2}{3}\left(e^{\frac{i2\pi}{3}}cos\xi\, e^{i\xi} + e^{-\frac{i2\pi}{3}}cos\eta\, e^{i\eta}\right)$ determined by two key parameters: $\xi = arctan\left[\frac{\omega - \omega_+}{\gamma}\right]$, $\eta = arctan\left[\frac{\omega - \omega_-}{\gamma}\right]$. $\gamma$ is the inverse decay time to ports (identical for all ports due to $C_3$ symmetry), and $\omega$ is the excitation frequency.

The symmetry and nonreciprocity can be concluded as follows: Reciprocal limits ($\xi = \eta$) occur when $\omega^+ = \omega^-$ (no magnetic field bias on the circulator), reducing $S_0$ to a symmetric matrix. Maximal nonreciprocity ($\xi = -\eta$) is achieved at $\omega = \frac{1}{2}(\omega^+ + \omega^-)$, where isolation (e.g., $|T| \neq |D|$) is maximized. Unitarity constraint: $S_0^\dagger S_0 = I_3$ enforces energy conservation, linking reflection ($|R|$) and transmission ($|T|$, $|D|$) magnitudes via $|R|^2 + |T|^2 + |D|^2 = 1$. The angular parameters $\xi, \eta$ map spectral detuning onto a compact parameter space, enabling topological phase transitions as shown in Fig. 1c.

## II. Derivation of the Bloch Eigenproblem for Nonreciprocal Scattering Networks

The Kagome graph like lattice comprises two identical $C_3$ symmetric nonreciprocal elements (denoted as 'A' and 'B' here) with oppositely oriented magnetic field biases (Fig. 1b, main text). In unit cell $n$, we define three state vectors as: $|a_n\rangle$ outgoing waves from all scatterer ports, $|b_n\rangle$ incoming waves to all scatterer ports, and $|c_n\rangle$ waves propagating along inter-element links. Notably, each vector here contains 6 complex amplitudes, thus leading to 6 bands within one Floquet spectral cycle.

As shown in Fig. S1, the scattering process at elements 'A' and 'B' is governed by the unitary operator

$$\boldsymbol{S}_{\text{elements}}|\boldsymbol{c}_n\rangle = \boldsymbol{P}_0 \begin{bmatrix} \boldsymbol{S}_A & 0 \\ 0 & \boldsymbol{S}_B \end{bmatrix} |\boldsymbol{c}_n\rangle = |\boldsymbol{a}_n\rangle \tag{S2}$$

where $\boldsymbol{P}_0 = U_{12} U_{13} U_{45} U_{56} \boldsymbol{I}_6$ is a $6 \times 6$ unitary permutation matrix enforcing connectivity, each $U_{ij}$ is an elementary permutation matrix swapping rows $i$ and $j$, $\boldsymbol{S}_A = \boldsymbol{S}_0(\xi, \eta)$, and $\boldsymbol{S}_B = \boldsymbol{S}_0(\eta, \xi)$. The periodicity of $\boldsymbol{S}_{\text{elements}}$ ensures translational invariance. Inter-element links impart a Bloch phase delay $\varphi$ and momentum-dependent phases via the unitary matrix:

$$\boldsymbol{\Lambda}(\boldsymbol{k}) = diag\left(e^{i\boldsymbol{k}\cdot\boldsymbol{\alpha}_2}, e^{i\boldsymbol{k}\cdot\boldsymbol{\alpha}_1}, 1, 1, e^{-i\boldsymbol{k}\cdot\boldsymbol{\alpha}_2}, e^{-i\boldsymbol{k}\cdot\boldsymbol{\alpha}_1}\right) \tag{S3}$$

where $\boldsymbol{\alpha}_1, \boldsymbol{\alpha}_2$ are Bravais lattice vectors. This relates outgoing and inter-link waves as $|\boldsymbol{a}(\boldsymbol{k})\rangle = e^{-i\varphi}\cdot\boldsymbol{\Lambda}(\boldsymbol{k})|\boldsymbol{c}(\boldsymbol{k})\rangle$.

Combining Eqs. (S2) and (S3) yields the central eigenvalue equation of a unitary Floquet operator with eigenproblems:

$$\boldsymbol{S}(\boldsymbol{k})|\boldsymbol{c}(\boldsymbol{k})\rangle = e^{-i\varphi}|\boldsymbol{c}(\boldsymbol{k})\rangle \tag{S4}$$

with $|\boldsymbol{c}(\boldsymbol{k})\rangle \in \mathbb{C}^6$ containing all inter-link wave amplitudes, and

$$\boldsymbol{S}(\boldsymbol{k}) \equiv \boldsymbol{\Lambda}^{-1}(\boldsymbol{k}) \boldsymbol{S}_{\text{elements}} = \boldsymbol{\Lambda}^{-1}(\boldsymbol{k}) \boldsymbol{P}_0 \begin{bmatrix} \boldsymbol{S}_A & 0 \\ 0 & \boldsymbol{S}_B \end{bmatrix} \tag{S5}$$

encapsulates both lattice geometry and nonreciprocal scattering. The real-valued quasi-energy $\varphi(\boldsymbol{k})$ arises from the unitarity of $\boldsymbol{S}(\boldsymbol{k})$.

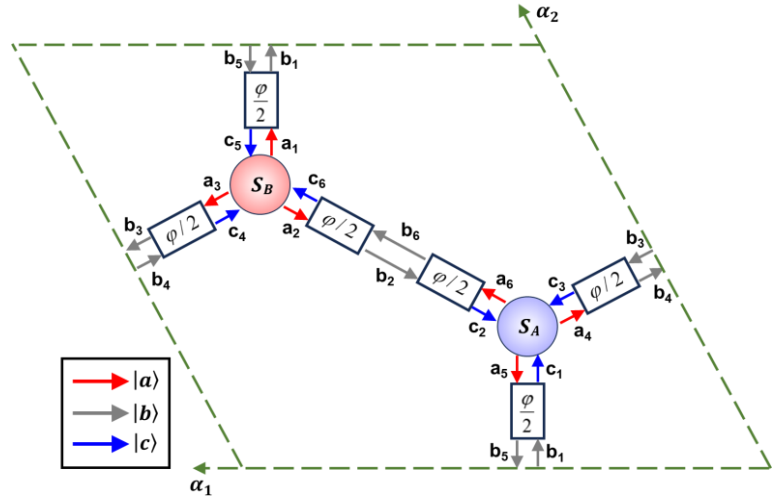

**Fig. S1 | Unit cell architecture and amplitude convention for a non-reciprocal network.** State vectors $|a_n\rangle$ (outgoing waves), $|b_n\rangle$ (intermediate propagation), and $|c_n\rangle$ (incoming waves) define scattering amplitudes. φ indicates total inter-element phase delay. $\boldsymbol{\alpha_1}, \boldsymbol{\alpha_2}$ are Bravais lattice vectors for a unit cell.

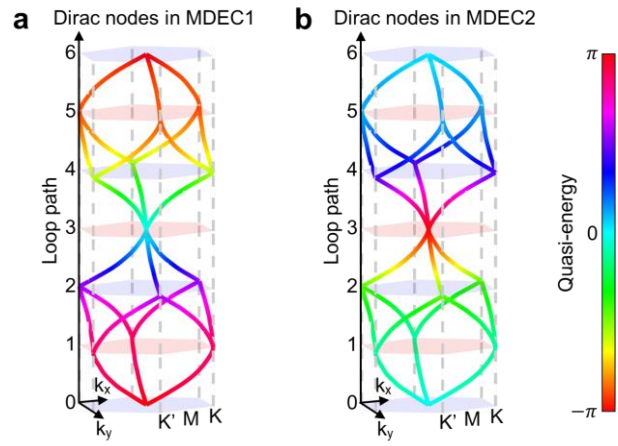

**Fig. S2 | Detailed Dirac nodes in MDEC1 (a) and MDEC2 (b) evolution along the loop path in Fig. 2e.** The distribution of node lines represents the momentum positions and the color denotes the quasi-energy. We note that the color finishes one cycle after rounding one cycle of loop path.

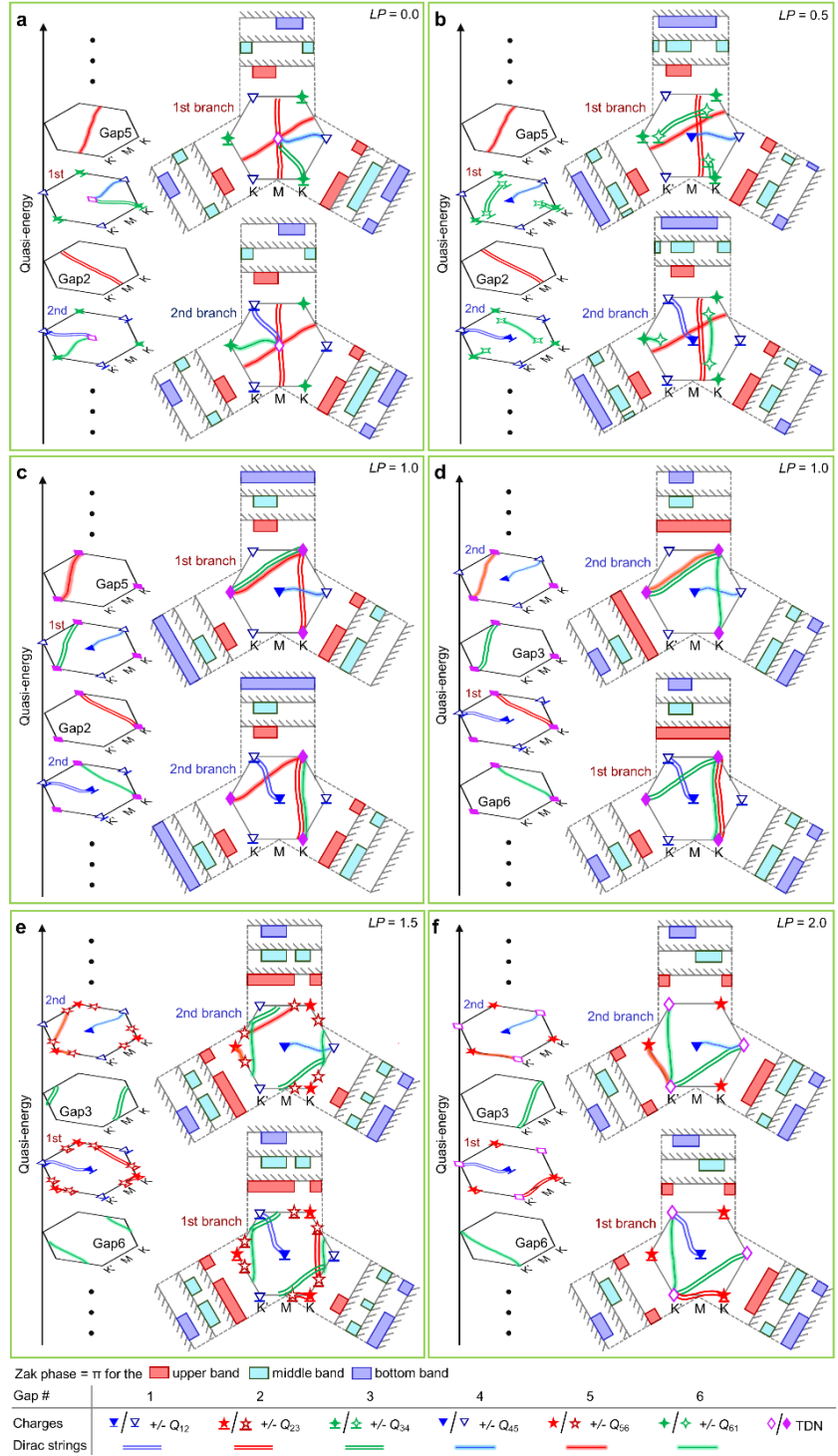

**Fig. S3 | Evolution of the Dirac strings and projected Zak phases along the loop path from 0 to 2. a-f**, The topology of the three zigzag edges is quantified by the non-trivial values of the Zak phase, which is integrated along high-symmetry paths traversing the Brillouin zone in the direction normal to the edges.

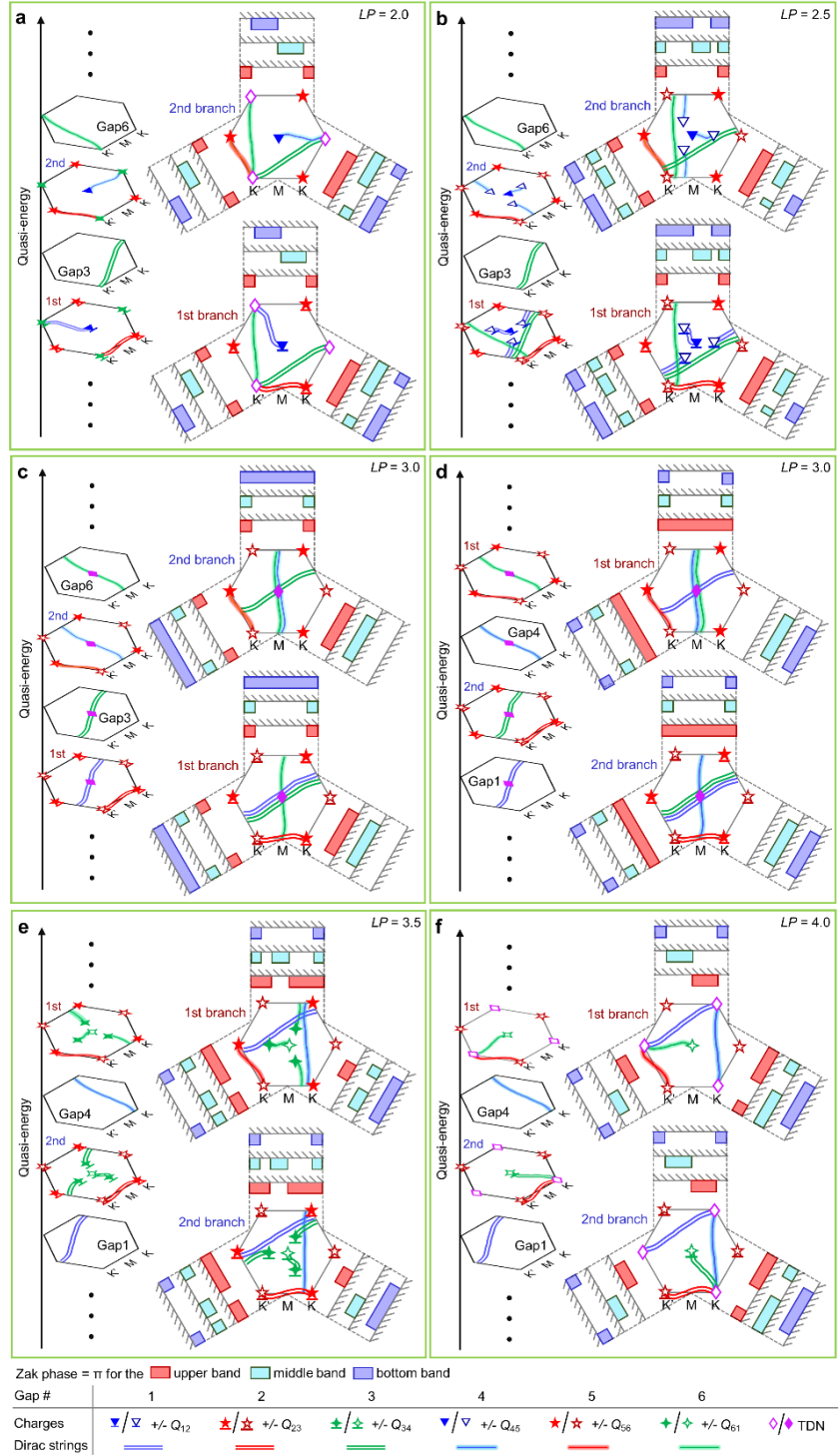

**Fig. S4 | Evolution of the Dirac strings and projected Zak phases along the loop path from 2 to 4. a-f**, The topology of the three zigzag edges is quantified by the non-trivial values of the Zak phase, which is integrated along high-symmetry paths traversing the Brillouin zone in the direction normal to the edges.

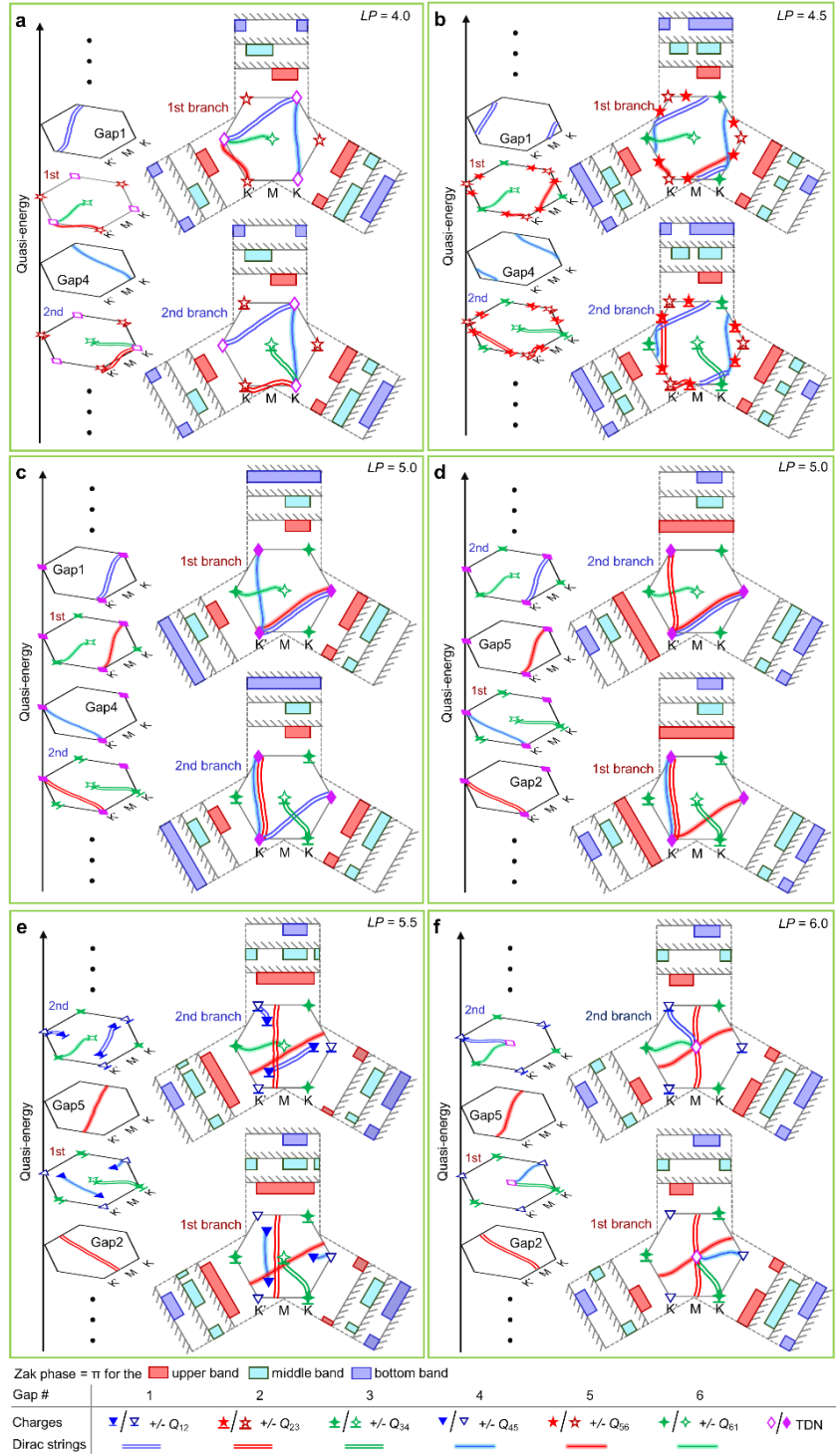

**Fig. S5 | Evolution of the Dirac strings and projected Zak phases along the loop path from 4 to 6. a-f**, The topology of the three zigzag edges is quantified by the non-trivial values of the Zak phase, which is integrated along high-symmetry paths traversing the Brillouin zone in the direction normal to the edges.

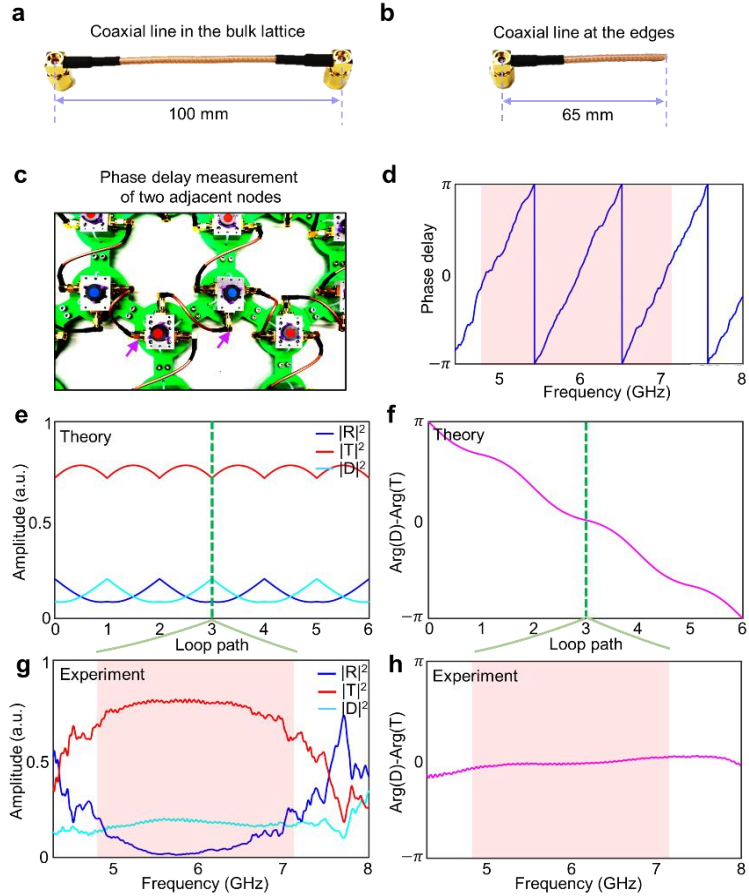

**Fig. S6 | Experimental network design and measured scattering parameters. a**, Coaxial transmission lines in the bulk lattice acting as links to impart frequency dependent phase delays between adjacent circulators. **b**, Coaxial transmission lines at *y*-direction edges are cut off to introduce boundary conditions with mirror reflection. **c**, A scheme to experimentally measure the phase delay of two adjacent notes in the lattice, and the measured nodes are marked by two arrows. **d**, Measured phase delay spectrum of two nodes in (c). The red-shaded area represents the bandwidth of around $4\pi$ phase covering, corresponding to two cycles of Floquet quasi-energy. **e**, Theoretical predictions of one scatter for CCW transmittance (T), CW transmittance (D) and reflection (R) spectra along the loop path in Fig. 2a. **f**, Theoretical prediction for the phase relation between two transmission spectra (CCW & CW). **g-h**, Experimental data of the scattering parameters corresponding to the theoretical results in (e-f). The red-shaded area containing the bandwidth of around $4\pi$ phase covering is very well matched with the theoretical spectra in the *LP* = 3.0, which shows the experimental feasibility of our proposed scattering network to validate the theoretical model prediction.